\documentclass[preprint2]{aastex62}

\def\beq{\begin{equation}}
\def\eeq{\end{equation}}

\def\aap{A\&A}
\def\apj{ApJ}

\def\mnras{MNRAS}
\def\apjl{ApJ}
\def\apjs{ApJS}

\def\prd{PRD}

\def\jcap{JCAP}
\def\memsai{Mem. Soc. Astron. Ital.}
\def\physrep{Phys. Rep.}

\begin{document}

\title{A Brief Review of Binary Driven Hypernova}

\author{Jorge~A.~Rueda}
\address{ICRA and Dipartimento di Fisica, Universit\`a  di Roma ``La Sapienza'', Piazzale Aldo Moro 5, I-00185 Roma, Italy \\
ICRANet, Piazza della Repubblica 10, I-65122 Pescara, Italy \\
ICRANet-Ferrara, Dipartimento di Fisica e Scienze della Terra, Universit\`a degli Studi di Ferrara, Via Saragat 1, I--44122 Ferrara, Italy \\
Dipartimento di Fisica e Scienze della Terra, Universit\`a degli Studi di Ferrara, Via Saragat 1, I--44122 Ferrara, Italy \\
INAF, Istituto di Astrofisica e Planetologia Spaziali, Via Fosso del Cavaliere 100, 00133 Rome, Italy \\
E-mail: jorge.rueda@icra.it}

\author{Remo~Ruffini}
\address{ICRA and Dipartimento di Fisica, Universit\`a  di Roma ``La Sapienza'', Piazzale Aldo Moro 5, I-00185 Roma, Italy \\
ICRANet, Piazza della Repubblica 10, I-65122 Pescara, Italy \\
INAF,Viale del Parco Mellini 84, 00136 Rome, Italy \\
E-mail: ruffini@icra.it}

\author{Rahim~Moradi and Yu~Wang$^*$}
\address{ICRA and Dipartimento di Fisica, Universit\`a  di Roma ``La Sapienza'', Piazzale Aldo Moro 5, I-00185 Roma, Italy \\
ICRANet, Piazza della Repubblica 10, I-65122 Pescara, Italy \\
INAF -- Osservatorio Astronomico d'Abruzzo,Via M. Maggini snc, I-64100, Teramo, Italy \\
E-mail: rahim.moradi@inaf.it \\
$^*$E-mail: yu.wang@inaf.it
}

\begin{abstract}
Binary driven hypernova (BdHN) models long gamma-ray burst (GRBs) as occurring in the binary systems involving a carbon-oxygen core (CO$_{\rm core}$) and a companion neutron star (NS) or a black hole (BH). This model, first proposed in 2012, succeeds and improves upon the fireshell model and the induced gravitational collapse (IGC) paradigm. After nearly a decade of development, the BdHN model has reached a nearly complete structure, explaining all the observables of long bursts into its theoretical framework, and has given a refined classification of long GRBs according to the original properties of the progenitors. In this article, we present a summary of the BdHN model and the physical processes at work in each of the envisaged Episodes during its occurrence and lifetime, duly contextualized in the framework of GRB observations.
\end{abstract}



\section{Introduction} 
\label{sec:intro}

GRBs occur in binary systems of two main classes, binary driven hypernovae (BdHNe) and binary mergers (BMs), observationally corresponding to long and short GRBs. In BdHNe, the long GRB is generated by a type Ic SN explosion of an evolved star occurring in presence of a close-by NS or BH companion.  This article is dedicated to the BdHN systems. In BMs, the short GRB is generated from the merger of two compact stars, mostly from the NS-NS systems, and for low energetic short GRBs, the binary white dwarfs could be the progenitors, see  \citet{2018JCAP...10..006R,2019JCAP...03..044R} and references therein.

\begin{table*}
\scriptsize
\centering
\begin{tabular}{c|cccccccc}
\hline
Class &   Type   & Number & \emph{In-state}  & \emph{Out-state} & $E_{\rm p,i}$ &  $E_{\rm iso}$  &  $E_{\rm iso,Gev}$  \\
& & & & & (MeV) & (erg) &  (erg) 	\\	
\hline
Binary Driven & I   & $329$ &CO$_{\rm core}$-NS  & $\nu$NS-BH & $\sim0.2$--$2$ &  $\sim 10^{52}$--$10^{54}$ &    $\gtrsim 10^{52}$ \\
Hypernova & II & $(30)$ &CO$_{\rm core}$-NS    & $\nu$NS-NS & $\sim 0.01$--$0.2$  &  $\sim 10^{50}$--$10^{52}$ &    $-$ \\
(BdHN) & III & $ (19) $ &CO$_{\rm core}$-NS    & $\nu$NS-NS & $\sim 0.01$  &  $\sim 10^{48}$--$10^{50}$ &    $-$ \\
& IV   & $5$ & CO$_{\rm core}$-BH  & $\nu$NS-BH & $\gtrsim2$ &  $>10^{54}$ &   $\gtrsim 10^{53}$   \\
\hline
\end{tabular}
\caption{Summary of the BdHN subclasses.  Values are taken from  \citep{2016ApJ...832..136R, 2016arXiv160203545R, 2017IJMPD..2630016R} with some updates. The ``number'' is the GRBs with known redshift identified of each subclass till the end of 2016 (bracket indicates the lower limit). The ``in-state'' and ``out-state'' represent the progenitors and outcomes. We also present the peak energy $ E_{\rm p,i}$, the isotropic gamma-ray energy, $E_{\rm iso}$ of $1$~keV to $10$~MeV energy range, and the isotropic ultra high energy $E_{\rm iso,Gev}$ of $0.1$--$100$ GeV energy range. This table is reproduced from \citep{2019ApJ...874...39W}.}
\label{tab:GRBsubclasses}
\end{table*}


The progenitor of a BdHN is a binary system composed of a carbon-oxygen core (CO$_{\rm core}$) and a magnetized neutron star (NS) companion in a tight orbit (period of the order of a few minutes). In some cases, the companion might be a stellar-mass BH (see below). We focus here on the more frequent case of an NS companion. At the end of its thermonuclear evolution, the iron core of the pre-SN star (the CO$_{\rm core}$) undergoes gravitational collapse, forming a new NS (hereafter $\nu$NS) at the SN centre. In the $\nu$NS formation process, a strong shockwave of kinetic energy $\sim 10^{51}$~erg expands outward and when it emerges (SN breakout) expels the CO$_{\rm core}$ outer layers as the SN ejecta. Part of the ejecta is subsequently accreted onto the companion NS and also onto the $\nu$NS by fallback. There are different possible fates for the NS due to the hypercritical accretion process \citep{2015ApJ...812..100B,2016ApJ...833..107B,2018ApJ...852..120B}. For short binary periods ($\lesssim 5$~min), the NS reaches the critical mass for gravitational collapse and forms a BH. We call this subclass BdHN of type I (BdHN I). Thus, a BdHN I leads to a new binary composed of a $\nu$NS originated by the SN, and a BH originated by the collapse of the NS companion. For longer binary periods, the hypercritical accretion onto the NS is not sufficient to bring it to the critical mass, and a more massive NS (MNS) is formed. This subclass is named BdHN of type II (BdHN II). A BdHN II leads to a new binary composed of a $\nu$NS and a massive NS. For very long binary periods, the accretion energy is significantly lower than the above types, and only the hypernova is observed. We call this subclass of sources of type III (BdHN III). In addition, we have BdHN type IV (BdHN IV) for the progenitors of a CO$_{\rm core}$ and a companion BH, and it leads to a new binary of NS and BH.

Having given the physical picture and the classifications of BdHNe, we will present in the following sections the theory and associated observables of the BdHNe. The BdHN starts from the final evolution of the binary stars, including the SN and the accretion of the SN ejecta onto the companion star, to the formation of BH and the particle acceleration mechanisms processing therein, then to the generated relativistic outflow propagates and interacts with the SN ejecta and the interstellar medium (ISM) giving rise to the emissions.

\section{Binary Accretion}

For the binary accretion and the forming of BH, we refer to the theoretical work of \citet{2012ApJ...758L...7R, 2014ApJ...793L..36F, 2015ApJ...812..100B, 2015PhRvL.115w1102F, 2016ApJ...833..107B, 2017PhRvD..96b4046C, 2018ApJ...852..120B, 2019ApJ...871...14B} and the observational papers of \citet{2012A&A...548L...5I, 2012A&A...543A..10I, 2015ApJ...798...10R, 2019ApJ...874...39W, 2021MNRAS.504.5301R}.

\begin{figure*}[!hbt]
\centering
\includegraphics[width=0.95\hsize]{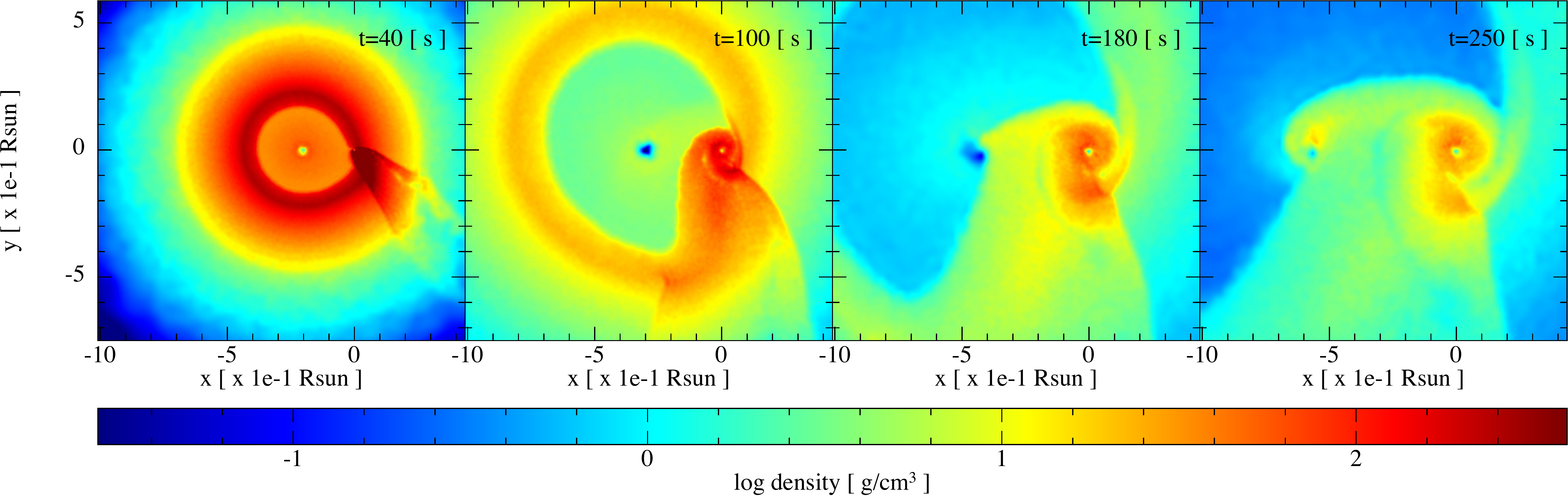}
\includegraphics[width=0.95\hsize]{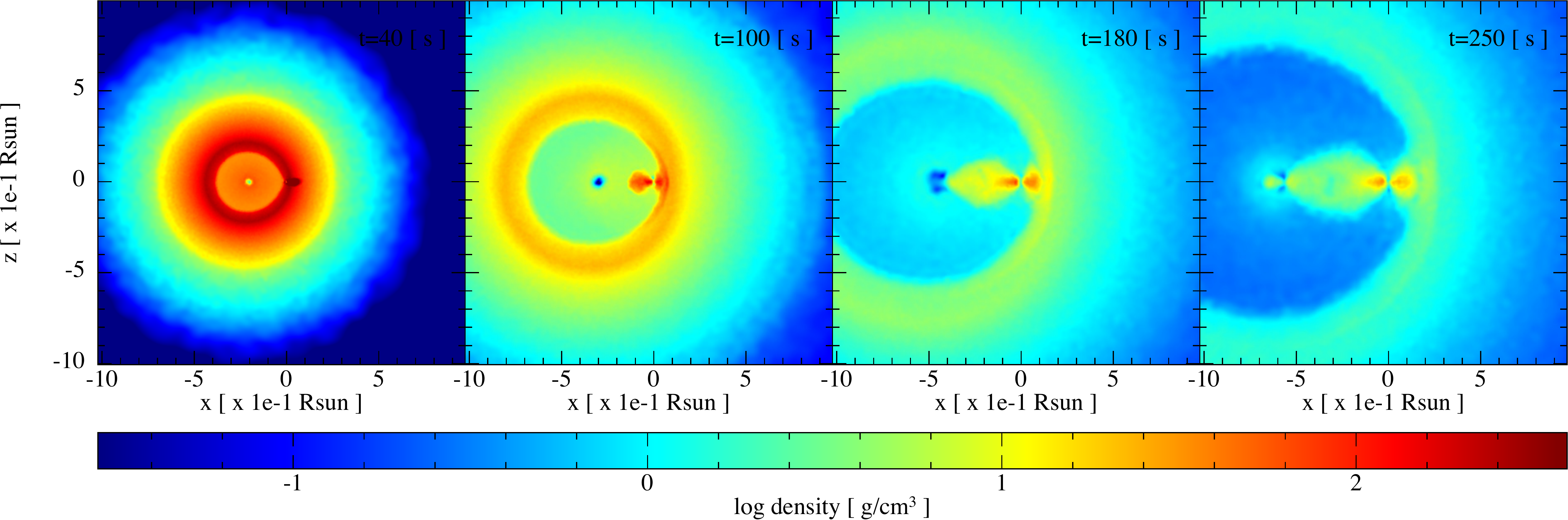}
\caption{Snapshot of the SPH simulation of the binary accretion. The initial binary system consists of a CO$_{rm core}$ ($M_{rm zams}=25 M_\odot$)  and an NS ($2 M_\odot$) with an initial orbital period of about $5$~min. The upper panel shows the mass density on the equatorial plane of the binary at different times of the simulation, while the lower panel corresponds to the plane orthogonal to the equator.  At $t=40$~s, particles captured by NS can be seen forming a kind of tail behind them, then these particles form a circle around NS, and at $t=100$~s a thick disk is observed. At $t=180$~s, NS starts to accrete the surrounding matters. After about one initial orbital period, at $t=250$~s, a disk-like structure has formed around the two stars. This figure is cited from \citet{2019ApJ...871...14B}.}
  \label{fig:simulation}
\end{figure*}

\begin{figure*}[!hbt]
\centering
\includegraphics[width=0.85\hsize,clip]{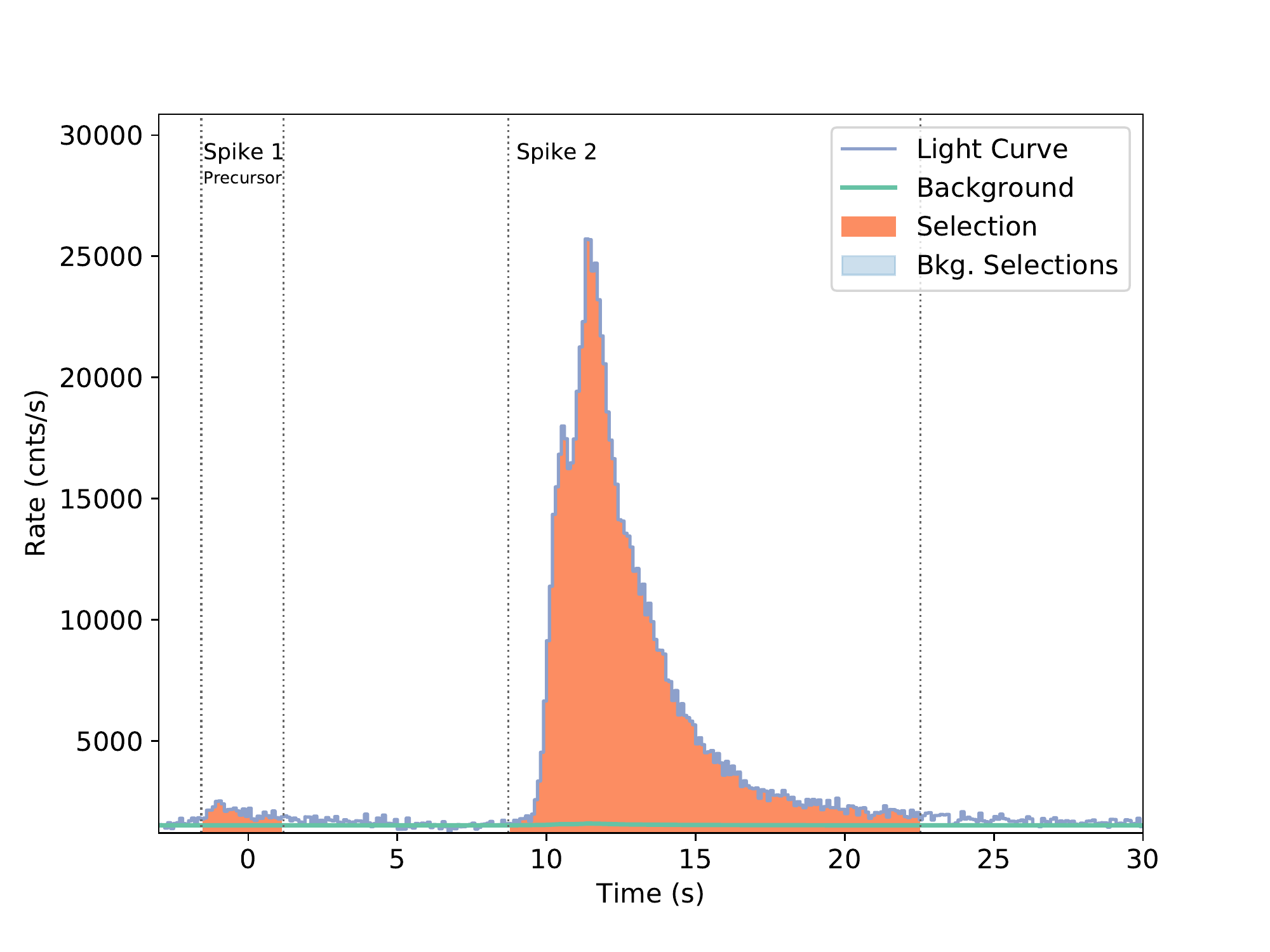}
\caption{Light curve of prompt emission of GRB 180728A observed by Fermi-GBM. It contains two pulses. The first pulse ranges from $-1.57$~s to$1.18$~s. The second pulse rises at $8.72$~s, peaks at $11.50$~s, and fades at $22.54$~s. This figure is quoted from \cite{2019ApJ...874...39W}.}
\label{fig:lightcurve}
\end{figure*}

\citet{2012ApJ...758L...7R} has been the first article to consider accretion of SN ejecta onto a very close-by companion star of the binary period of minutes, and it gave the physical picture and the theoretical architectures of a simple one-dimensional model that calculates the Bondi-Hoyle-Lyttleton hypercritical accretion rate. \citet{2014ApJ...793L..36F} numerically simulated for the first time the BdHN hypercritical accretion. Following the collapse of the $CO_{\rm core}$ of forming an SN, of which the ejecta falling onto the Bondi-Hoyle surface of the companion with an accretion rate $>10^{-2} \rm M_\odot~\rm s^{-1}$, these one-dimensional numerical simulations give the density and the velocity profiles till the NS reaches the critical mass of BH in tens or hundreds of seconds. \citet{2015ApJ...812..100B} went one step further performing two-dimensional numerical simulations and incorporating angular momentum transport from the SN ejecta to the NS of hypercritical accretion. Those simulations show that under some conditions outflow is necessarily formed because of the excess of angular momentum. \citet{2015PhRvL.115w1102F} demonstrated that most BdHN with tight orbits (i.e. BdHN I) remain bound after the explosion and accretion, even when a large fraction of mass (over half of the total binary mass) is lost, and a large kick velocity is induced. \citet{2016ApJ...833..107B} performed the first three-dimensional numerical simulations of the BdHNe process, which were further upgraded and improved in \citet{2019ApJ...871...14B} via smoothed-particle-hydrodynamics (SPH) simulations, see e.g. in figure \ref{fig:simulation}. A wide selection of initial parameters and several NS equations of state have been there tested. It was there evaluated the outcome of the NS and the $\nu$NS after the hypercritical accretion, namely whether they reach or not the mass-shedding limit, or gravitationally collapses to a BH, or become a more massive and fast-spinning NS. The development of accretion theory and simulations has led to clarifying the physical processes below the Bondi radius: the dominant pressure is supported by the random pressure of the infalling matter, the magnetic pressure is negligible. Such a pressure provides a very high temperature of $1$--$10$ MeV, generating a large abundance of neutrinos and photons. The photons are trapped within the inflowing material, as its diffusion velocity is slower than the inflow velocity. The escape of neutrinos takes away most gravitational energy from the accreted flow, allowing the hypercritical accretion to continue for a given period. \citet{2018ApJ...852..120B} further investigated the neutrino flavour oscillations that occur during the propagation of neutrinos emitted from the surface of a neutron star. The final neutrino flow is composed of $\sim55\%$ ($\sim$62\%) of $\sim$ MeV electronic neutrinos for the normal (inverted) neutrino mass hierarchy. In addition, \citet{2017PhRvD..96b4046C} present the numerical calculation and give useful fitting formulas for the location, binding energy and angular momentum of the last stable orbit of test particles around the rotating NSs in full general relativity. The results of this work allow estimating in full general relativity the amount of energy and angular transferred by the accreting matter to an accreting, rotating NS.

Let us now dive into one specific example. GRB 180728A well demonstrates the binary accretion scenario by its two pulses in the prompt emission \citep{2019ApJ...874...39W}, see figure \ref{fig:lightcurve}. At a given time, the CO$_{\rm core}$ collapses, forms a $\nu$NS, and produces an SN explosion. A powerful shockwave is generated and emerges from the SN ejecta. A typical SN shockwave carries $\sim 10^{51}$~erg of kinetic energy, which is partially converted into electromagnetic radiation with an efficiency of $\sim 10\%$. Thus, the energy of $\sim 10^{50}$~erg is consistent with the total energy of the first pulse, which lasts $\sim 2$~s and contains $\sim 8 \times 10^{49}$~erg in keV-MeV photons. The second pulse rises at $\sim 10$~s and subsides at $\sim 10$~s, with a luminosity $\sim 2 \times 51$ erg s$^{-1}$. The distance of the binary separation can be estimated by the delay time between these two pulses, i.e. $\sim 10$~s. Because the SN ejecta front shell moves at $\sim 0.1 c$, we estimate the distance of the binary separation to be about $3\times 10^{10}$~cm. By given the binary separation and some typical initial parameters, our simulation shows the total mass accreted is $\sim 10^{-2}~M_{\odot}$, most of the mass is accreted in $\sim 10$~s with an accretion rate of $\sim 10^{-3}~M_{\odot}$~s$^{-1}$. These results are consistent with the second pulse whose total energy is $\sim 10^{51}$~erg, considering an increase in efficiency of $\sim 10\%$, and the luminosity of $\sim 10^{50}$~erg~s$^{-1}$ in $10$~s duration. The spectrum of the second pulse contains a thermal component which again hints at the action of the accretion process. A time-resolved analysis of the thermal component suggests that a mildly relativistic source is expanding and radiating. This radiation is interpreted as an adiabatic expansion heat outflow from the accretion region. The Rayleigh-Taylor convective instability plays a role in the initial accretion phase, driving matter out of the accreting NS with a final velocity of the order of the speed of light. As the matter expands and cools, the temperature evolution from the theory is again consistent with the observations. This kind of thermal emission of BdHN was first found in \cite{2012A&A...548L...5I, 2012A&A...543A..10I}. GRB 180728A offers a good example of BdHN II, and for BdHN I, which has a tighter binary separation, see also the case of GRB 130427A as an example \citep{2015ApJ...798...10R, 2019ApJ...874...39W}. 


\section{Inner Engine}

We have introduced the inner engine theory for the explanation of the GRB high-energy (GeV) emission observed in some BdHN I after the prompt emission phase. We here summarize the inner engine properties following the calculations presented in \citet{2019ApJ...886...82R, 2020EPJC...80..300R, 2021A&A...649A..75M, CAMPION2021136562}; see also \citet{2018arXiv181101839R, 2019arXiv190404162R, 2019arXiv191012615L, 2021arXiv210309158M, 2021arXiv2108}.

\begin{figure*}[!htb]
\centering
\includegraphics[width=0.4\hsize,clip]{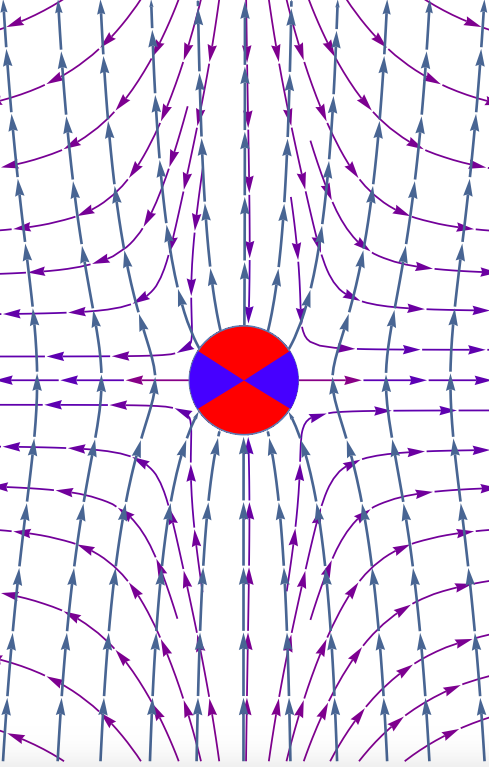}
\vline \vline \vline \vline \vline \vline \vline
\includegraphics[width=0.4\hsize,clip]{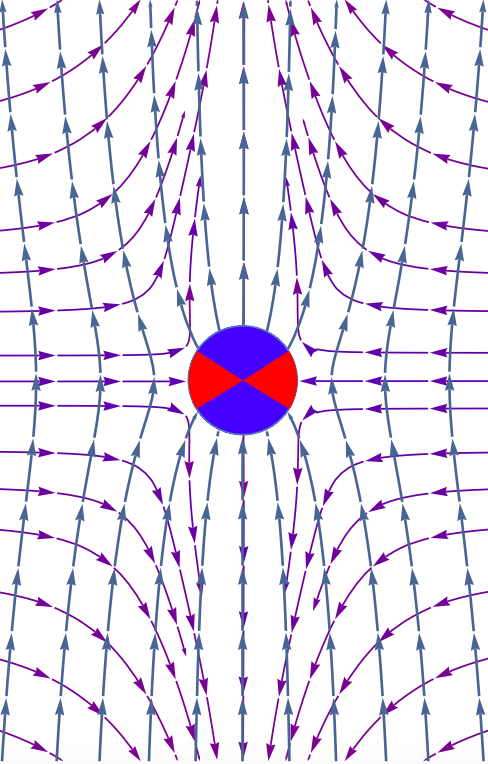}
\caption{The electromagnetic field lines of Wald solution. The blue lines indicate the magnetic field lines, and the purple lines indicate the electric field lines. \textbf{left:}. The magnetic field is parallel to the spin of the Kerr BH, so parallel to the rotation axis. The electric field lines are inward for the polar angle $ \theta < \sim\pi/3$, so the electrons are accelerated away from the BH. For $\theta > \pi/3$, the electric field lines are outward, so the protons are accelerated away from the BH. \textbf{right:} The magnetic field is antiparallel to the rotational axis of the Kerr BH. The electric field lines are outward for the polar angle $\theta < \sim\pi/3$, so the protons will be accelerated away from the BH. The electric field lines are inward for $\theta > \pi/3 $, so the electrons will be accelerated away from the BH. This figure is quoted from \citet{2019ApJ...886...82R}.}
\label{fig:field}
\end{figure*}

\begin{figure*}[!htb]
\centering
\includegraphics[width=1\hsize,clip]{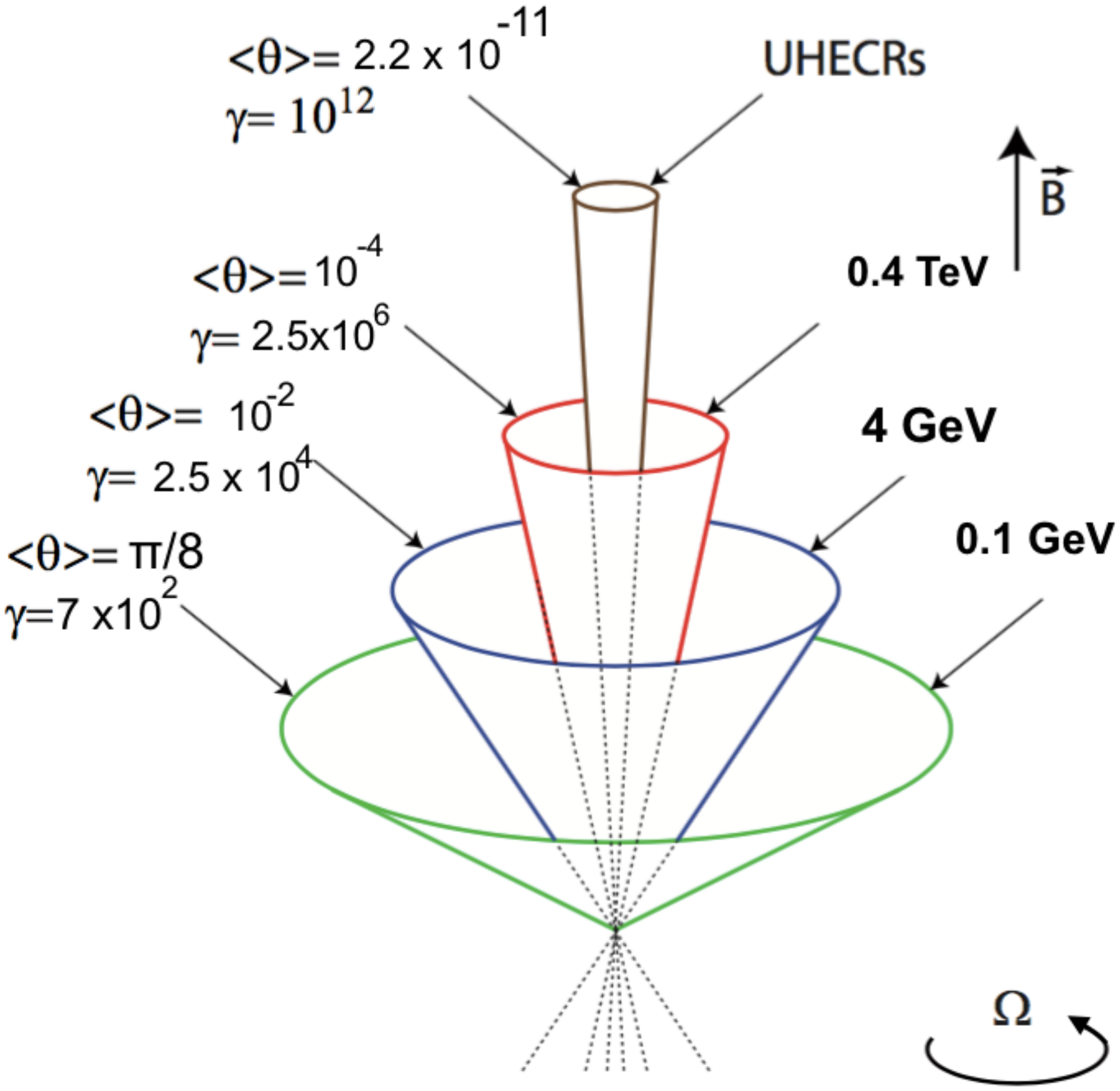}
\caption{The radiation emitted by the synchrotron emission of accelerated electrons in the different bands ($0.1$ GeV to $0.4$ TeV) from different angles. And in the narrow polar cone, The UHECRs are produced. Arrows indicate the BH rotational direction and the external magnetic field direction.  This figure is quoted from \citet{2019ApJ...886...82R}.}
\label{fig:cones}%
\end{figure*}

Once the NS reaches the critical mass, a fast-rotating BH forms which contain sufficient rotational energy ($> 10^{54}$~erg) to power a GRB.  \citet{2019ApJ...886...82R} proposed an efficient way to extract energy from the newborn, Kerr BH. 

The inner engine is composed of this newborn rotating BH, surrounded by the magnetic field inherited from the collapsed NS \citep{2020ApJ...893..148R}, and the ionized very low density of matter ($\sim 10^{-14}$~g~cm$^{-3}$) of the SN ejecta \citep{2019ApJ...883..191R}. For an aligned magnetic field with the angular momentum of the Kerr BH, an electric field is induced by gravitomagnetic interaction as described by the Wald solution \citep{Wald:1974np}; see figure \ref{fig:field}. The medium around the BH provides a sufficient amount of ionized particles that are accelerated to ultra-relativistic energies by the induced electric field, thereby emitting synchrotron and curvature radiation at the expense of the BH rotation energy.

The synchrotron radiation emitted from the accelerated charged particles has been calculated for different polar angles, see figure \ref{fig:cones}. Along the polar axis, the electric and magnetic fields are aligned, so there are no radiation losses and electrons can reach energies as large as $\sim 10^{18}$~eV, becoming a source of ultrahigh-energy cosmic rays (UHECRs). At larger angles, where electrons propagate across the magnetic lines producing synchrotron photons in the GeV energy domain.

The parameters of the inner engine, namely the BH mass and spin, and the surrounding magnetic field strength have been inferred from the following conditions: 1) the Kerr BH extractable energy accounts for the observed GeV radiation energetics in BdHN I, 2) the synchrotron radiation luminosity explains the observed GeV luminosity, and 3) the emitted GeV photons can indeed scape from the system without suffering from magnetic pair production. The case of GRB 130427A has been analyzed in \citet{2019ApJ...886...82R} and GRB 190114C in \citet{2021A&A...649A..75M}. For instance, in GRB 190114C the accelerated electrons radiate $1.8 \times 10^{53}$~erg in GeV photons via the synchrotron mechanism, and this procedure lasts for years following a power-law decay of light-curve with power-law index $-1.2$.

The mass and spin of the BHs came precisely what was expected from the gravitational collapse of the fast rotating NS by accretion, and the strength of the magnetic field surrounding the BH turns out to be a few $10^{10}$~G. Therefore, the intensity of the induced accelerating electric field is undercritical in this BdHN I episode. The above magnetic field strength is lower than expected to be inherited from the NS, and that could be the result of a screening process during the GRB prompt emission by electron-positron pairs; see the next section and \citet{CAMPION2021136562} for details on this interesting physical process. In a recent comprehensive analysis of all the up-to-know identified BdHN I and the GeV emission observed in some of them, it has been inferred that the GeV emission must be emitted within an angle of $60^\circ$ from the BH rotation axis; see \citet{2021MNRAS.504.5301R} for details. This result is in agreement with the theoretical expectation; see \citet{2021A&A...649A..75M}.

Another discovery of this inner engine model is that this energetic emission from GRB is not continuous but proceeds in a repetitive sequence of discrete impulse events. Since the inner engine repeats the procedure of charge (BH spin and magnetic field induce electric field) and discharge (electric field accelerates the charged particle that escapes from the system), the repetition time grows slowly along with the loss of BH rotational energy. Along with the emission of these discrete events, the magnetic field keeps constant, but the BH spin decreases after each event by a well-defined amount given by the concept of \textit{blackholic quantum} described in \citet{2020EPJC...80..300R}. The blackholic quanta explaining the GeV emission are characterized by an energy $\Delta E_q\sim 10^{38}$~erg, emitted over a timescale $\tau_q \sim 10^{-15}$~s. The fraction of BH angular momentum extracted after each event is $\Delta J_q/J \sim 10^{-16}$, i.e. $\Delta J_q\sim 10^{33}$ g cm$^2$ s$^{-1}$ \citep{2021A&A...649A..75M}, where $J$ is the Kerr BH angular momentum. This result is indeed unexpected, and it seems to be a general property not only of GRBs but also of the supermassive Kerr BHs in active galactic nuclei; see e.g. \citet{2021A&A...649A..75M} for the analysis of M87*. 

\section{Prompt and Afterglow Emissions}

This section gives the details of the propagation and the radiation of the relativistic outflow, based on the theoretical articles of \citet{1999A&A...350..334R, 2000A&A...359..855R, 2018ApJ...869..101R, 2010PhR...487....1R} and the observational articles of \citet{2015ApJ...798...10R, 2015ARep...59..667W, 2018ApJ...852...53R, 2018MmSAI..89..293W, 2018ApJ...869..151R, 2019ApJ...874...39W, 2020ApJ...893..148R}.

\begin{figure*}[!htb]
\centering
\includegraphics[width=0.9\hsize,clip]{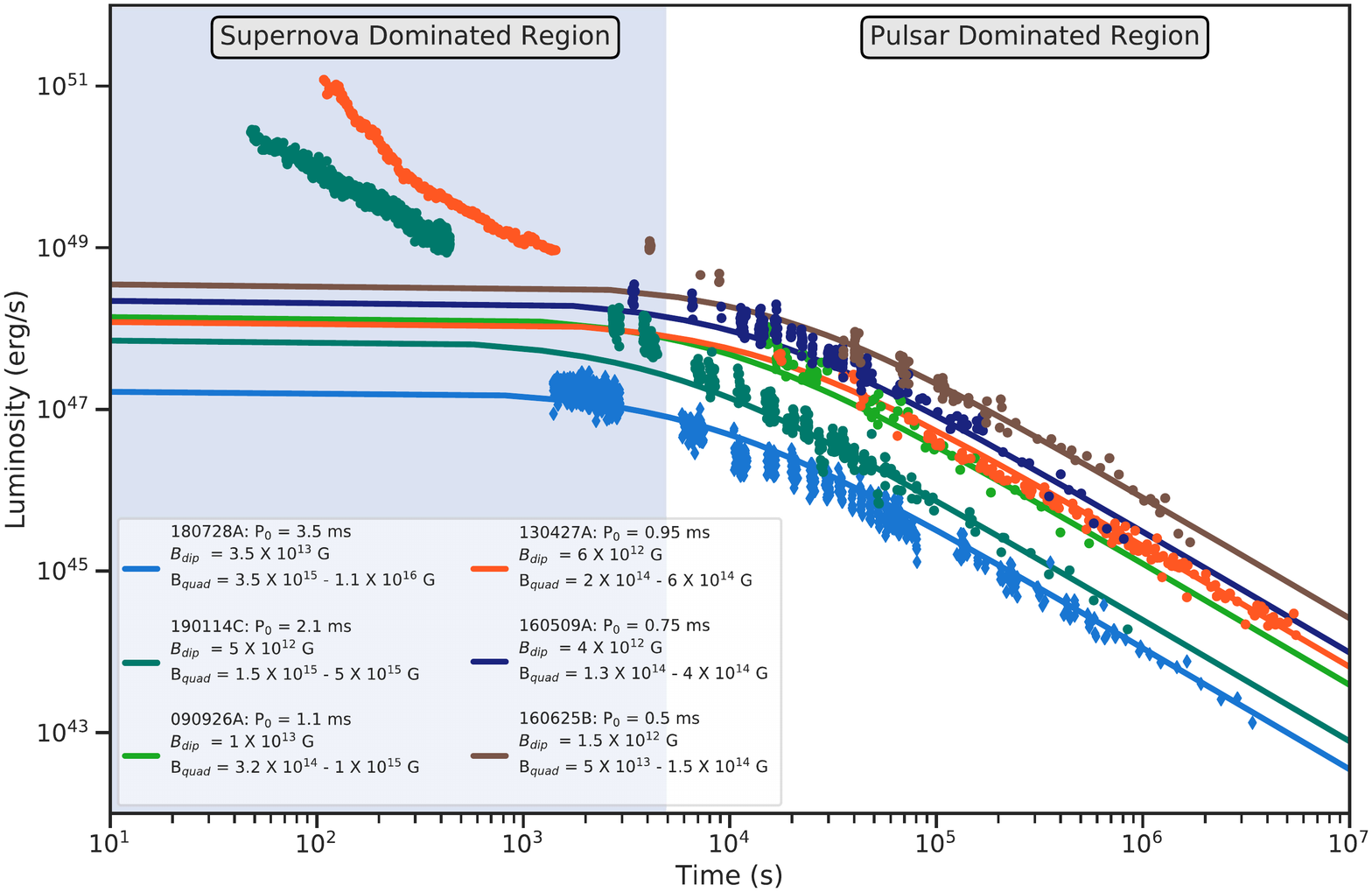}
\caption{The brown, dark blue, orange, green, and bright blue dots correspond to the bolometric light curves of GRB 160625B, 160509A, 130427A, 190114C, and 180728A, respectively. The thick lines are energy injection from the NS spin-down, which energizes the late-time afterglow (white background), while in the early time (blue background) the remaining kinetic energy of the outermost shell of the SN ejecta plays a dominant role. This figure is quoted from \citet{2020ApJ...893..148R}.}
\label{fig:lightcurves}
\end{figure*}

\subsection{Prompt Emission Phase}

\citet{2021arXiv2108} investigated the ultra-relativistic prompt emission (UPE) phase of the BdHN I in which the electric field is overcritical, generating an optically thick electron-positron plasma by vacuum polarization. The plasma expands and self-accelerates to ultra-relativistic by converting its internal energy. Eventually, it reaches the transparency point and releases photons in the MeV energy domain. \citet{CAMPION2021136562} studied the channel of producing electron-positron pairs via high energy photons interacting with magnetic fields, the motion of these pairs generates a current which induces another dominant magnetic field that screens the original one.

We now focus on the relativistic electron-position plasma that leads to the UPE phase. The hydrodynamics of this plasma has been formulated and simulated in the articles that established the \textit{fireshell} model, and which have been adopted by the BdHN model. \citet{1999A&A...350..334R} considered a Reissner-Nordstrom electromagnetic BH generates the electron-positron pairs. The system is expected to be thermalized to a plasma configuration due to the huge pair density and cross-section of the $e^{+}+e^{-} \rightarrow \gamma + \gamma$ process. The evolution of plasma is governed by the hydrodynamic equations including, the conservation of energy-momentum, the conservation of the baryon number, the rate equation for electron-positron annihilation, and the equation of state. By integrating the equations numerically under the Reissner-Nordstr\"om metric and compared with the analytical analysis, the temperature drops as the internal energy are converted to kinetic energy. The plasma Lorentz factor is accelerated to several hundred, and the radiation releases when reaching the transparency radius. \citet{2000A&A...359..855R} adopted a more realistic baryonic environment where the plasma propagates. Baryons are incorporated into the hydrodynamic equations. It is clear by solving the equations that the Lorentz factor of plasma keeps increasing in the beginning until engulfing the baryons, a drop of Lorentz factor occurs, then it goes up again and finally reaches saturation. \citet{2010PhR...487....1R} systematically reviewed the creation and annihilation of the electron-positron pairs, thermalization, oscillation and their applications in GRB observations. The plasma penetrates the low-density region of the SN ejecta, see \citet{2016ApJ...833..107B} and \citet{2018ApJ...852...53R} for details, then propagates in the ISM, and radiates, accounting for the prompt emission. For the observations, the light-curve and spectrum of the prompt emission have been successfully fitted by solving the hydrodynamic equations plus a density profile of the circumburst medium. \citet{2012A&A...543A..10I} offers an example of GRB 090618, of which the system starts by $\sim 2.5 \times 10^{53}$~erg electron-position plasma. The plasma propagates in the circumburst density of $0.6$ cm$^{-3}$, and collides with dense clouds of mass $\sim 10^{24}$~g at the distance of $10^{15}$~cm to $10^{16}$~cm. The plasma finally self-accelerates up to transparency reaching a Lorentz factor $\sim 500$, thereby producing the observed emission. 

We refer the reader to the most recent analysis of the UPE phase in BdHN I presented in \citet{2021arXiv2108}, where the physical origin of the UPE phase has been scrutinized taking as a proxy GRB 190114C. The UPE phase of GRB 190114C is observed in the rest-frame time interval $t_{\rm rf}=1.9$--$3.99$~s, by the \textit{Fermi}-GBM in $10$ keV--$10$ MeV energy band. Thanks to the high signal–to–noise ratio of \textit{Fermi}-GBM data, time-resolved spectral analysis of the UPE emission has evidenced a sequence of similar blackbody plus cut-off power-law spectra (BB+CPL), on ever decreasing time intervals. In it, the inner engine operates in an overcritical electric field regime. The electron-positron pair electromagnetic plasma in presence of a baryon load, \emph{a PEMB pulse}, is therein originated from a vacuum polarization quantum process. This initially optically thick plasma self-accelerates, giving rise at the transparency radius to the MeV radiation observed by \textit{Fermi}-GBM. For the first time, it has been quantitatively shown how the inner engine, by extracting the rotational energy of the Kerr BH, produces a series of PEMB pulses. Therefore, a quantum vacuum polarization process sequence with decreasing time bins occurs. We compute the Lorentz factors, the baryon loads and the radii at transparency, as well as the value of the magnetic field in each sequence. It has been therefore found there is an underlying fundamental hierarchical structure, linking the quantum electrodynamics regime of the UPE to the classical electrodynamics regime of the GeV emission after the UPE. The PEMB pulses of the UPE are characterized by the emission of blackholic quanta of energy $\sim 10^{45}$~erg, over a timescale $\sim 10^{-9}$~s. 

Let us summarize GRB 190114C. The initial magnetic field left over by the collapse of the accreting NS and rooted in the surrounding material is very strong ($\sim 10^{14}$~G), so it induces a sizeable electric field that surpasses the critical value near the horizon. The overcritical electric field transfers its energy to the electron-positron pairs by the vacuum polarization and is later emitted as the UPE phase of $2.5 \times 10^{53}$~erg. The magnetic field becomes then screened to a few $10^{10}$~G in a few seconds \citep{CAMPION2021136562}, consequently, the size of the region above the BH horizon with an overcritical electric field (the \textit{dyadoregion}) shrinks, and its energy stored becomes insignificant. This marks the end of the UPE phase and after it, the above inner engine mechanism by which the induced electric field accelerates electrons within a few horizon radii becomes the main channel of taking away the BH rotation energy in form of GeV photons \citep{2021A&A...649A..75M}.

\subsection{Afterglow Emission Phase}

Another part of the plasma hindered by the SN ejecta accelerates the SN outermost shell to mildly-relativistic velocities. The breakout of the plasma (shockwave) from the outermost shell at $\sim 10^2$~s radiate photons of keV energies which explain the observed X-ray flares \citep{2018ApJ...852...53R}. The synchrotron emission in the outermost shell accounts for the early afterglow X-ray emission \citep{2015ApJ...798...10R}. Rotational energy from the $\nu$NS rotational is injected into the SN ejecta, then radiated by the synchrotron emission, accounts for the plateau and late-time ($\sim 10^4$~s) afterglow \citep{2018ApJ...869..101R, 2019ApJ...874...39W, 2020ApJ...893..148R}. The emission of the $\nu$NS as a pulsar becomes directly observable when the synchrotron luminosity fades off below the pulsar radiation luminosity. About $\sim 15$ days (rest-frame time) after the SN explosion that triggered the BdHN, the optical emission from the nickel decay in the SN ejecta reaches the maximum, there may appear a bump on the optical light-curve \citep{2015ApJ...798...10R, 2015ARep...59..667W, 2019ApJ...874...39W}.

\citet{2018ApJ...852...53R, 2018MmSAI..89..293W} statistically analyzed the X-ray flares observed in the early afterglow. A general pattern of thermal component of temperature $\sim 1$~keV was found, suggesting that the flare is generated from a mildly-relativistic expanding shell of Lorentz factor $< 4$ at a distance $\sim 10^{12}$~cm. The observation of flares is consistent with our simulation of $\sim ~10^{53}$ erg of plasma impacts on the SN ejecta of a few solar masses, leading to the formation of a shock propagating inside the SN ejecta until reaching the outermost shell. The density profile and velocity profile of the accelerated ejecta are obtained. Precisely, the shockwave breaks out at $\sim 10^{12}$~cm and the outermost shell is accelerated to Lorentz factor $\sim 2$--$5$. This feature was also extensively studied for GRB 151027A \citep{2018ApJ...869..151R}. Along with the conversion of the kinetic energy of the outermost shell into radiation, the early afterglow exhibits a steep decay behaviour. Then, the energy injected from the $\nu$NS dominates the afterglow, the light-curve shows a plateau followed by a normal power-law decay, shown in figure \ref{fig:lightcurves}. Taking GRB 180728A as an example \citep{2019ApJ...874...39W}, from the conversion of angular momentum, the CO$_{\rm core}$ collapses to a fast spinning NS of initial spin period $\sim 3$~ms. Such a newborn NS allows the presence of multipolar magnetic fields, with a quadrupole magnetic ﬁeld $\sim 10^{15}$~G and a dipole field $\sim 10^{13}$~G, the spin-down of the $\nu$NS injects energy into the outflowing ejecta whose synchrotron emission fits the late-time X-ray afterglow. A comprehensive analysis of the afterglow of a few long GRB afterglows within the above synchrotron mechanism of the BdHN model has been presented in \citet{2020ApJ...893..148R}.

\section{Conclusion}

\begin{figure*}[htb]
    \centering
    \includegraphics[width=0.9\hsize]{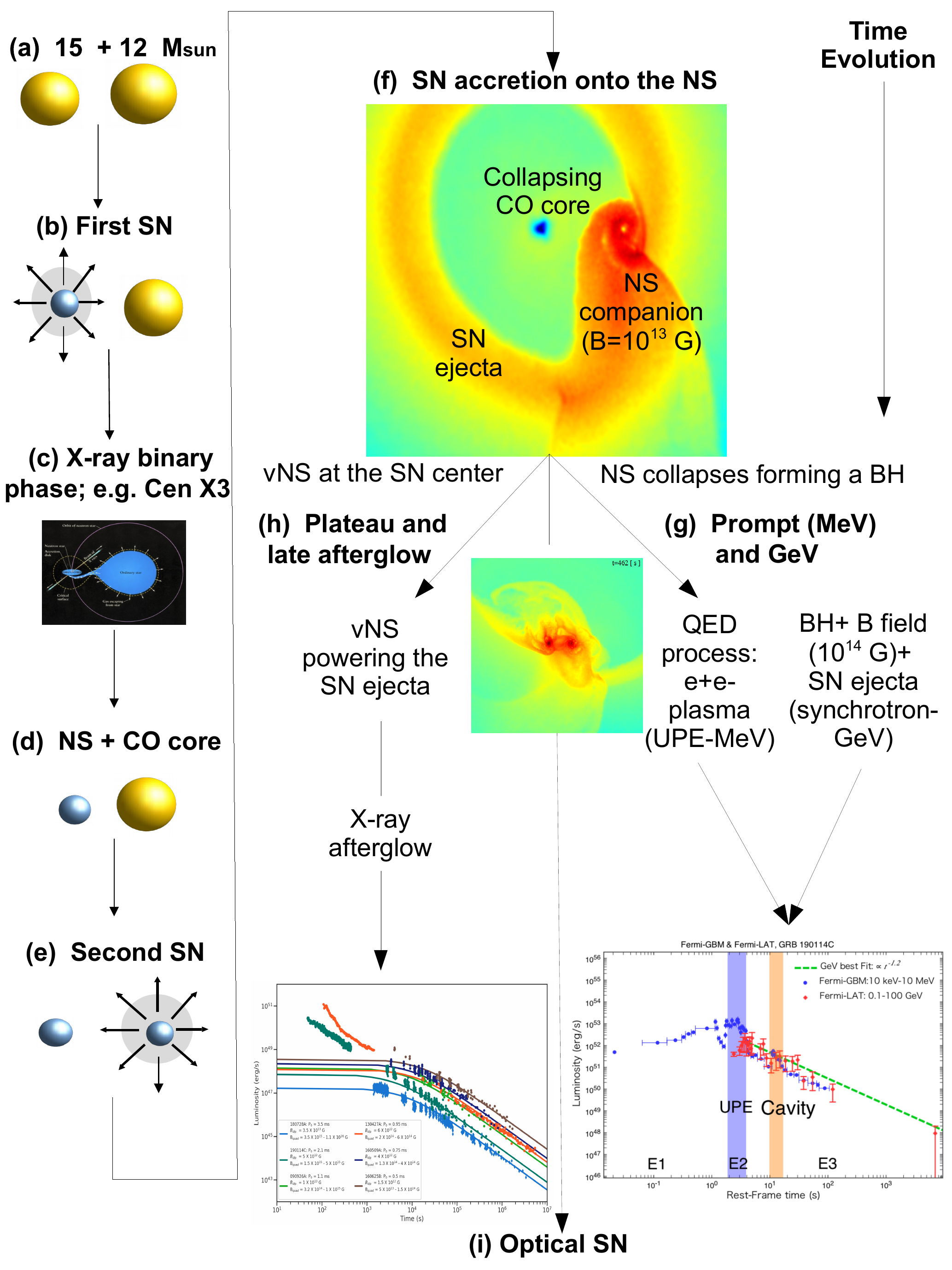}
    \caption{Diagram of the evolutionary path of BdHN. Including binary evolution, SN explosion, NS accretion, BH formation, GRB prompt and afterglow emissions and SN appearence. This figure is quoted from \cite{2020ApJ...893..148R}.}
    \label{fig:BdHN}
\end{figure*}

We can draw some general conclusions with the aid of the BdHN evolution shown in figure \ref{fig:BdHN}. (a) Our picture starts with a binary system consisting of two main-sequence stars of intermediate mass, say $15 M_\odot$ and $12 M_\odot$, respectively. (b) At a given time, the more massive star undergoes a core-collapse SN and forms an NS. (c) The system enters the X-ray binary phase. (d) The system has overcome binary interactions and common-envelope phases (not shown in the diagram) which have led to the hydrogen and helium envelopes of the ordinary star to have been stripped off, remaining a star which is rich in carbon and oxygen, referred to as CO$_{\rm core}$. At this stage, the system is a CO$_{\rm core}$-NS binary, which is considered as the initial configuration of the BdHN model. (e) At this stage, the orbit of the binary has shrunk to a period of the order of a few minutes. The CO$_{\rm core}$ explodes into an SN (of type Ic in view of the absence of hydrogen and helium in its outermost layers), expelling several solar masses. This ejecta begins to expand, and a rapidly rotating $\nu$NS is left in the centre. (f)  Depending on the initial NS mass and binary separation, the SN ejecta accretes onto the NS companion and the $\nu$NS, forming a massive NS (BdHN II) or BH (BdHN I; this example). At this stage, the system is a new NS and BH binary surrounded by the expanding ejecta. (g) The inner engine composed of the newborn BH, the surrounding magnetic field and ionized plasma is formed, and its activity explains the GRB UPE phase and the subsequent GeV emission. The magnetic field in the inner engine at BH formation is overcritical, so it induces (by gravitomagnetic interaction with the BH spin) an overcritical electric field, so the UPE phase operates in an overcritical regime. A quantum electrodynamical process of vacuum polarization takes place leading to electron-positron pair plasma pulses (PEMB pulses) that expand to ultrarelativistic velocity reaching transparency with a Lorentz factor of up to hundreds, and emitting MeV photons. The magnetic field is then screened to undercritical values by currents produced by the motion of the electron and positrons, and the inner engine classical electrodynamical process of particle acceleration emitting GeV photons by synchrotron radiation becomes the relevant process of emission. (h) The spin-down energy of the $\nu$NS injects energy into the expanding SN ejecta emitting the observed X-ray afterglow by synchrotron radiation. (i) The appearance of the energy release owing to nickel decay in the SN ejecta is observed at optical wavelengths.

\section*{Acknowledgments}

We acknowledge the contribution of the ICRANet group and the collaborators in establishing the BdHN model. Jorge Rueda, Remo Ruffini, and Yu Wang gave the plenary talks in the MG 15 meeting on behalf of the contributors.

\end{document}